# Correcting METIS spectra for telluric absorption to maximize spectral fidelity

Stefan Uttenthaler[*a], Klaus M. Pontoppidan[b], Andreas Seifahrt[c], Sarah Kendrew[d], Joris A. D. L. Blommaert[a], Eric J. Pantin[e], Bernhard R. Brandl[d], Frank J. Molster[f], Lars Venema[g], Rainer Lenzen[h], Phil Parr-Burman[i], Ralf Siebenmorgen[j], and the METIS team

[a]Instituut voor Sterrenkunde, K. U. Leuven, Celestijnenlaan 200D, 3001 Leuven, Belgium; [b]California Institute of Technology, Division of Geological and Planetary Sciences, Pasadena, CA 91125, USA; [c]University of California, Physics Department, One Shields Avenue, Davis, CA 95616, USA; [d]Leiden Observatory, University of Leiden, P. O. Box 9513, 2300 RA Leiden, The Netherlands; [e]CE Saclay DSM/IRFU/SAp, Service d'Astrophysique, Batiment 709, l'Orme Les Merisiers, 91191 Gif sur Yvette Cedex, France; [f]NOVA, P. O. Box 9513, 2300 RA Leiden, The Netherlands; [g]NOVA/ASTRON, Oude Hoogeveensedijk 4, 7991 PD Dwingeloo, The Netherlands; [h]Max-Planck-Instiut für Astronomie, Königsstuhl 17, 69117 Heidelberg, Germany; [i]UK Astronomy Technology Centre, Royal Observatory Edinburgh, Blackford Hill, Edinburgh EH9 3HJ, UK; [j]ESO Garching, Karl-Schwarzschild-Strasse 2, 85748 Garching, Germany; [*]stefan@ster.kuleuven.be

**ABSTRACT**

METIS is a mid-infrared instrument proposed for the European Extremely Large Telescope (E-ELT). It is designed to provide imaging and spectroscopic capabilities in the 3 – 14 μm region up to a spectral resolution of $10^5$. One of the novel concepts of METIS is that of a high-resolution integral field spectrograph (IFS) for a diffraction-limited mid-IR instrument. While this concept has many scientific and operational advantages over a long-slit spectrograph, one drawback is that the spectral resolution changes over the field of view. This has an impact on the procedures to correct for telluric absorption lines imprinted on the science spectra. They are a major obstacle in the quest to maximize spectral fidelity, the ability to distinguish a weak spectral feature from the continuum. The classical technique of division by a standard star spectrum, observed in a single IFS spaxel, cannot simply be applied to all spaxels, because the spectral resolution changes from spaxel to spaxel. Here we present and discuss possible techniques of telluric line correction of METIS IFS spectra, including the application of synthetic model spectra of telluric transmission, to maximize spectral fidelity.

## 1. INTRODUCTION

METIS is the name of the "**M**id-infrared **ELT I**mager and **S**pectrograph", the only instrument in the phase-A study for the European Extremely Large Telescope (E-ELT) to cover the thermal/mid-infrared (MIR) wavelength range from 3 – 14 μm. Complementary to observational capabilities that are or will be provided by ongoing IR space projects like Herschel and the James Webb Space Telescope, METIS will concentrate on high spatial and spectral resolution: All observation modes are diffraction limited reaching 17 milli-arcseconds resolution at 3.5 μm, which is quite similar to the high-resolution mode of the Atacama Large Millimeter Array. This spatial resolution can be combined with a spectral resolving power of up to R = $\lambda/\Delta\lambda$ = 100,000. Based on the METIS science case presented elsewhere at this conference[1], the following two main sub-systems have been designed.

1) A diffraction-limited **imager** in the atmospheric L-, M-, and N-band windows with an approximately 18"×18" field of view (FOV), including the following sub-modes:

- L/M- and N-band broad and narrow-band imaging
- L/M- and N-band low-resolution, long slit spectroscopy (R = 300 to 3000)
- L/M- and N-band coronography
- N-band linear polarimetry

2) An integral field unit (IFU) fed **high-resolution spectrograph** at L/M-band (2.9 – 5.3 μm.) The IFU FOV will be approximately 0.4"×1.5", and the spectral resolution R = 100,000. Details of the system engineering and the optical design of METIS are presented elsewhere at this conference[2].

## 2. SPECTRAL FIDELITY

The concept of a high-resolution integral field spectrograph (IFS) for a diffraction limited MIR instrument is a novelty. While this concept has many scientific and operational advantages over a long-slit spectrograph[1], one drawback is that the spectral resolution changes over the field of view such that even neighboring spaxels can have a quite different spectral resolution. This is a purely geometrical effect caused by the varying angles at which the rays hit the grating surface. The effect is the same for an IFS and a long-slit spectrograph, and depends only on the (resorted) slit length. This has an impact on the calibration procedures of the spectra, in particular on the correction of the telluric lines. They are usually divided out of the science spectrum by a feature-less standard star spectrum observed close in time and at similar airmass. However, this requires that both spectra have exactly the same resolution – or the resulting spectrum will be dominated by residuals that do not depend on the photon flux and may limit the capabilities to detect faint lines well above the theoretical sensitivity of the instrument.

The relevant quantity here is the spectral fidelity. Spectral fidelity describes the ability to distinguish a weak spectral feature from the surrounding continuum: A fidelity of 1:100 refers to a feature of 1% of the continuum flux that can still be distinguished from the continuum (signal-to-continuum). It is different from the signal-to-noise ratio (SNR) as it does not depend on the photon flux. There are several METIS science cases that require a spectral fidelity of 1:1000 or better. Most importantly, these requirements come from the detection of faint gas lines of protoplanetary disks[13], the detection of planetary spectral signatures in exoplanet transit observations, and the direct detection of disk line emission from circum*planetary* disks. The results with existing instrumentation have underlined the importance of getting the telluric correction right.

Assuming a Gaussian instrumental line profile (ILP), the relation between the strength of the residual $\Psi$ and the relative change in spectral resolution can be described by:

$$\Psi \approx \left(1 - e^{-\tau}\right)\left(\left(1 + \frac{\Delta R}{R}\right)^2 - 1\right),$$

where $\tau$ is the optical depth of the (unresolved) telluric line. Assuming $\tau = 1$ for a non-saturated telluric line, we can rewrite the above equation and get:

$$\frac{\Delta R}{R} \approx \sqrt{1.582\,\Psi + 1} - 1 \,.$$

Hence, a fidelity of 1:100 corresponds to $\Delta R/R \sim 0.008$ and a fidelity of 1:1000 to $\Delta R/R \sim 0.0008$. In other words, a fidelity requirement of 1:1000, together with a requirement on the spectral resolution of R = 100,000, would fix the actual spectral resolution across the FOV to within 99,960 and 100,040. As discussed above, due to geometrical effects within the main dispersion element, this is not possible to meet within acceptable cost limits. Broad water lines will be resolved at this resolution, but e.g. the very narrow ozone lines with very little pressure broadening are not fully resolved. Since it would be extremely time consuming and impractical to place a telluric standard star in every IFU field point ("spaxel") every time a spectrum of an extended source is taken, other possibilities to correct the science spectrum for telluric line absorption and to maximize its spectral fidelity have to be found. The aim of this paper is to explore different possibilities to achieve this goal for METIS science observations.

# 3. CORRECTING METIS SPECTRA FOR TELLURIC LINES

In the following, we will assume that the spectra have been properly calibrated otherwise, i.e. dark subtracted, sky subtracted, flat fielded, etc. (wavelength calibration is not necessarily required). If these calibrations have been applied, we are then left with three basic possibilities for correcting the telluric absorption lines in the spectrum:

1. Division by a telluric standard star spectrum, observed close in time to the science spectrum.
2. Division by a synthetic model spectrum of the telluric transmission.
3. A combination of those two.

Which one of these techniques will lead to the better result depends mainly on the spectral characteristics of the science target. If it has many intrinsic features or is not a source of continuum radiation, the classical technique of division by a standard star spectrum will probably yield higher spectral fidelity, because it will be very difficult to find a satisfying telluric transmission model by fitting the science spectrum itself. Also, fitting the science spectrum directly will not be possible if it has a very low SNR. However, if the science target has only few intrinsic features (say, one absorption line that is of interest), then a telluric absorption model can be fitted directly to its spectrum, and a division will give the unperturbed spectrum.

# 4. MODEL SPECTRA OF ATMOSPHERIC TRANSMISSION

The method involving synthetic spectra of the Earth's atmosphere transmission requires a model of the atmospheric structure (temperature, pressure, and molecular densities as a function of altitude) and a list of (molecular) line transitions. With the help of a radiative transfer code, a synthetic spectrum of the transmission can be calculated. Some parameters such as water vapor content, ILP (resolution), wavelength calibration, but also the temperature structure of the atmosphere itself may be fitted to the observed spectrum. A division of the observed spectrum by this synthetic spectrum will, in the ideal case, retrieve the spectrum of the object as it would have been observed in the absence of the Earth's atmosphere.

The use of a telluric transmission model has a number of advantages over the classical approach of division by a standard star spectrum: i) The atmospheric properties such as seeing (which influences the spectral resolution), airmass, water content of the atmosphere, temperature structure, etc., can be modeled exactly for the time and direction of the science observations. ii) No further noise will be added to the final spectrum. iii) The science spectrum can be wavelength calibrated very precisely with the telluric lines, the precision will basically only be limited by the velocity of high atmospheric winds, which can be of the order of tens of meters per second[3]. Of course the observed spectrum can be calibrated independently of the model using the laboratory wavelengths of a few telluric lines, but this solution can be further improved during the fitting process. iv) The achieved resolution and the ILP can be derived directly from the science spectrum, if enough narrow lines are present in the observed spectrum. Figure 1 demonstrates that at least for the CRyogenic high-resolution InfraRed Echelle Spectrograph (CRIRES[4]) the ILP deviates very little from a Gaussian function. At the moment, it is not quantified which spectral ranges contain enough unresolved lines to determine the ILP, and which not. v) Finally, it will save expensive observing time if fewer or no standard star observations have to be carried out, making more night time available for science observations.

There are, however, also a number of disadvantages of using model spectra for telluric correction. i) If there are not narrow enough lines present in the observed wavelength range, the ILP cannot be directly determined from the science observations. Unless very high precision is required, a generic Gaussian profile may be assumed, which will generally give good results. To get an improved handle on the deviations from a Gaussian profile, a numeric kernel that has been pre-defined by other measurements, e.g. with a laser line or gas cell lines from the calibration unit, can be used. ii) There might be residual instrumental effects that are not sufficiently corrected for by the flat fielding process (due to e.g. light path and slit illumination differing between flat fielding source and real target) that will also not be corrected for by the model spectrum. iii) If the science spectrum is too complex or has a low SNR, it will be impossible to fit a model spectrum of telluric transmission directly to the science spectrum. iv) Finally, the available molecular line data (such as from the HITRAN 2008 data base[5]) might not be flawless at the level of accuracy required.

Analysis of CRIRES observations in selected wavelength settings[6] demonstrates that in the current 2008 edition of HITRAN, particularly the Q branches and transitions that have been extrapolated to high rotational quantum numbers (J) suffer from inaccuracies. Also, lines that exhibit a strong damping are not well described by a Voigt profile, which is in use in current codes to calculate synthetic transmission spectra (see Section 4.1). These inaccuracies cannot generally be constrained to certain wavelength ranges or molecular species, though $CO_2$ lines seem to be most problematic. The time available until first light of METIS, which is expected roughly in a decade from now, should be used to improve on this.

We recommend to work in collaboration with the HITRAN consortium to improve the molecular line data in the spectral range accessible with METIS. For this end, the existing line list needs to be tested with available high-resolution (e.g. CRIRES) spectra, and inaccuracies reported to the HITRAN team, to improve on the residual errors in the data. A complete assessment of the quality of the current molecular line list in the wavelength range accessible to METIS will be accomplished within the CRIRES-POP project[7]. This effort will have a high legacy value for all ground based IR instrumentation.

What needs to be available to METIS users in the astronomical community in order to be able to apply model spectra for the telluric correction? First of all, an (improved) HITRAN list and a radiative transfer code to calculate transmission spectra (see next section) need to be available. Also, an atmospheric model structure needs to be available. In the ideal case, a specific model for the site and the particular observing night or even the precise time of science observation is created. A collaboration on this between ESO and the meteorology group of the Vina del Mar University has already been established. Most important in such a model is the temperature structure and the amount and distribution of water vapor in the atmosphere. Also, the amount of other trace species such as $CO_2$ and $O_3$ vary with time and need to be constrained. The amount of precipitable water vapor needs to be continuously monitored during the METIS observations with the aid of a well-calibrated radiometer. This measurement will yield a starting value for the amount of water to be used in the model calculation that can be refined while fitting the spectra themselves, if suitable water lines are present in the observed wavelength range. Ideally, all of these tools are bundled into a single, efficient fitting code, and made available to the user as part of the METIS data reduction pipeline in order to take full advantage of the science spectra.

Figure 2 shows a demonstration of the level of accuracy that can be reached in the correction of telluric absorption lines by using model spectra. The standard deviation of the residuals, denoted in the lower panel by the dotted horizontal lines, is well below the 1% level of the continuum flux. However, the figure also demonstrates a draw-back of telluric transmission models: At the red end of this piece of spectrum, two water lines appear with significantly positive residuals. These two lines obviously cannot be fit simultaneously with a unique amount of water vapor in the atmosphere, or with the adopted water vapor distribution in the Earth's atmosphere. Possibly, at least one of them has wrong data in the HITRAN line list. Taking into account the cosmetic deficiencies of the CRIRES detector chip as well as the deficiencies in the line list, the residuals can be lowered to 0.5% of the continuum flux level, to reach a spectral fidelity of 200. This is in agreement with the intrinsic SNR of the spectrum. This is approximately the spectral fidelity that was reached in tests with existing CRIRES data. Note that these data have not been taken with the goal to reach a spectral fidelity of 1000, so a fidelity of 1:200 is certainly much below the limit of what can be achieved with current instruments and methods.

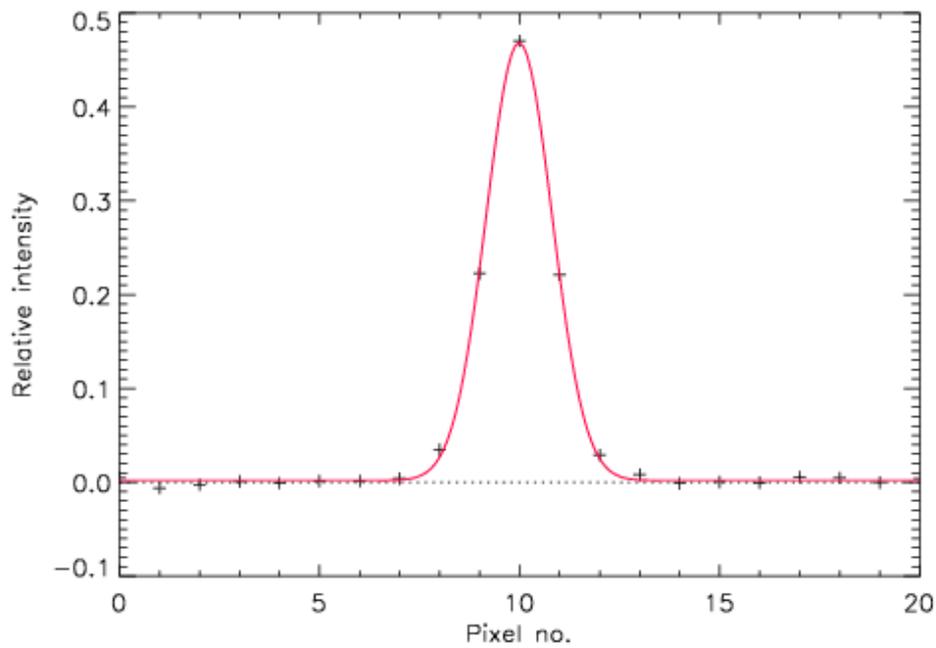

**Fig. 1:** Instrumental line profile of CRIRES (black crosses) determined from a standard star observation in the M-band that includes narrow $O_3$ lines. The red graph is a Gaussian fit to the profile. The deviation of the ILP from a Gaussian shape is very small. The profile has been determined by applying an SVD method[8], using a synthetic spectrum of telluric absorption as a sharp lined template spectrum.

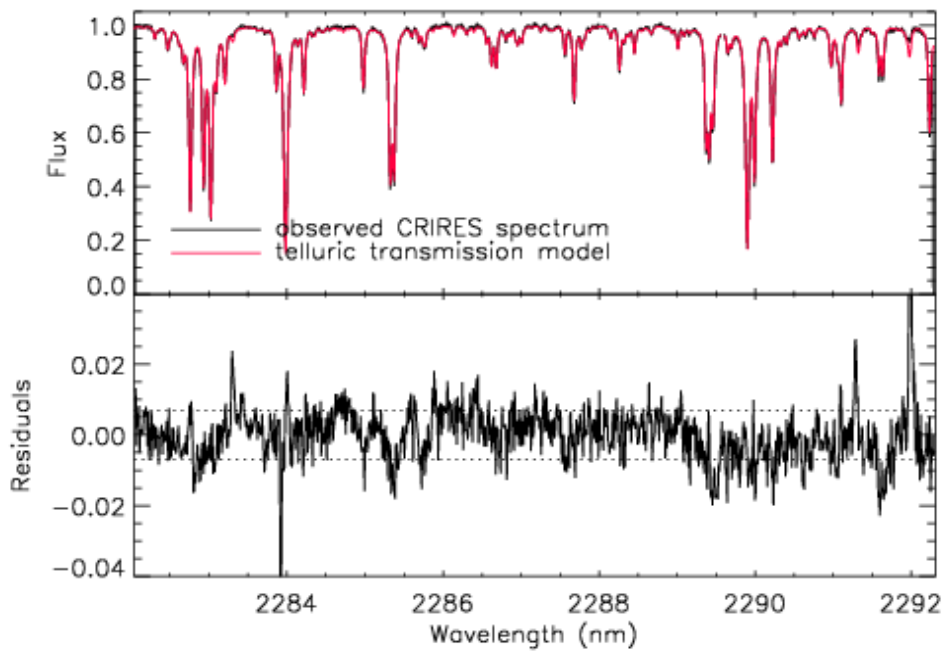

**Fig. 2:** Fit of a model of telluric transmission to an observed CRIRES spectrum. The upper panel shows the flux of an early-type star as observed with CRIRES (black graph), and a fitted RFM model of telluric transmission (red graph). The lower panel displays the flux difference between the two spectra. The dotted lines in the lower panel denote the ±1-sigma standard deviation of the residuals.

Besides fitting the telluric *absorption* lines in the science spectrum itself, another option is to model the sky *emission* spectrum. The sky emission spectrum can be obtained by observing a relatively empty region of the sky next to the science target. It is possible to constrain the ILP (e.g. from narrow OH emission lines), the pressure and temperature structure of the atmosphere, and the molecular column densities from such a model fit[10]. External constraints, e.g. from a radiometer, may be useful. A transmission spectrum of the such established atmospheric model can be calculated, and applied to correct the telluric absorption lines in the science spectrum. One draw-back of this method is that the such determined ILP is only useful for extended science targets. For point sources, the transmission model spectrum will have to be convolved with a generic Gaussian profile. As of yet, little experience with this method exists, but it should be considered to be applied to METIS observations. After all, many science targets might have too complex or too low-SNR spectra to fit a transmission model directly, hence such a calibration could be very useful. If observations are carried out with the imager of METIS, one interesting option is to monitor the atmospheric conditions throughout the night by observing empty sky regions in a suitable setting of the high-resolution IFS in parallel mode with the imager, without an additional investment of observing time.

**4.1 Comparison between two codes for calculating telluric transmission: LBLRTM and RFM**

In the course of this study we also made a comparison between two widely used and free of charge codes to calculate, amongst others, synthetic transmission spectra of the Earth's atmosphere: The Reference Forward Model (RFM, http://www.atm.ox.ac.uk/RFM/) and the Line-By-Line Radiative Transfer Model (LBLRTM, http://rtweb.aer.com/lblrtm.html). It has been proven recently[6,10] that both codes can be wrapped into a fitting algorithm to derive the atmospheric content of water and other molecular species, and use them to correct observed science spectra. We made sure that for both codes we used the same input parameters. We adopted the HITRAN08 molecular database[5], and used a merged atmospheric model that consisted of a Global Data Assimilation System (GDAS) model for Cerro Paranal on 10 August 2008 for the lower 20 km, and the ngt.atm MIPAS profile (http://www.atm.ox.ac.uk/RFM/atm/) constructed by John J. Remedios of the University of Leicester for the upper atmosphere between 20 and 120 km altitude. We calculated synthetic transmission spectra with both codes in the range 0.9 – 5.5 μm, assuming a location on Cerro Paranal (2.635 km altitude), and zenith viewing conditions. For RFM, a resolution of 2 Million was assumed; for LBLRTM, the resolution is determined by the code itself, such that even the narrowest lines in the investigated spectral range is sufficiently resolved. The spectra where smoothed to a resolution of 100,000 before comparing them. This smoothing was done externally from the codes using a the same method for both spectra.

Though the resulting spectra are very similar, we did detect some differences. First, the ratios of the spectra around reasonably strong lines ($\tau > 1$) showed asymmetric deviations from 1.0. This implies that probably one of the codes has rounding errors in the wavelengths of the molecular transitions that could become apparent at a high level of accuracy. RFM and LBLRTM also seem to have differences in the calculation of the continuum. This becomes particularly apparent in the wavelength range 4.12 – 4.19 μm, where the spectra of the two codes differ significantly. Comparison with an observed transmission spectrum[9] shows that the LBLRTM model spectrum is more realistic in this range. Finally, the treatment of $CO_2$ line mixing is problematic and does not lead to the same results in both codes, as the difference between the RFM and LBLRTM becomes quite large around $CO_2$ bands at 4.695, 4.775, 4.815, and 5.175 μm. These issues need to be investigated to improve the usage of synthetic model spectra of telluric transmission for the correction of astronomical science spectra.

# 5. STANDARD STAR OBSERVATIONS FOR TELLURIC CORRECTION

If the science spectrum is too complex (e.g. in the case of late-type stars or an emission spectrum without continuum) or has too low an SNR to unambiguously fit a model of telluric transmission to it, rather the use of standard star observations is recommended. Two standard star observations bracketing the science observations can make sure that the variation of water content between science and standard observations is kept to a minimum. If such a standard star spectrum is available, the following procedure might be followed to achieve a good correction:

1. Wavelength calibrate the science and the standard star spectrum independently, e.g. by using telluric lines. Since the telluric lines are imprinted on the spectra themselves, this allows for accurate calibration. In the science spectrum, telluric lines will be blended with intrinsic features, thus complicating their measurement.
2. Interpolate the standard star spectrum to the wavelength points of the science spectrum.
3. Extrapolate the standard star spectrum to the airmass of the science spectrum using $f_{std,extrapolated} = \exp\left(\ln(f_{std}) \times X_{sci}/X_{std}\right)$, where $X_{sci}$ and $X_{std}$ are the airmass at which the science target and the standard star were observed, and $f_{std}$ the flux of the standard star with continuum normalized to 1.0. This works fine for unsaturated absorption lines, and not too large differences in airmass.
4. Divide the science spectrum by the extrapolated and re-binned standard star spectrum. A division in the wavelength space rather than pixel space, as performed in this procedure, will reduce residuals considerably.

An advantage of using telluric standard star spectra is that the accuracy of molecular line data will of course not be an issue. Assuming that the seeing and Adaptive Optics (AO) performance are constant, the ILP will be the same for science and standard spectrum. Since this is not necessarily granted, an understanding of how the AO performance affects the spectrum is required to adapt the observational strategies. This is a task to be worked on during commissioning. If the ILP is indeed essentially the same for both observations, instrumental artifacts will be effectively corrected for.

However, there are also a number of disadvantages. i) Precious night time is "lost" to calibration observations. ii) The atmospheric conditions at the time and in the direction of the standard star observation might be different from those of the science observations. For instance, if the water content of the atmosphere changes in the meantime, residuals in the final spectrum around the (broad) water lines will remain. This cannot simply be corrected by extrapolating to the airmass of the science target. iii) According to Gauss' law of error propagation, noise will be added to the final spectrum by division by a (noisy) standard star spectrum. iv) As pointed out by ref. [11], the classical technique of division by a standard star spectrum is fundamentally flawed if the telluric lines are not fully resolved, and the intrinsic spectrum that would have been observed in the absence of the Earth's atmosphere cannot be restored. Note that this is a major driver for a high spectral resolution of the instrument, since at low resolution many telluric lines are not resolved[12]. It is particularly difficult to restore the original spectrum if the observed target has the same intrinsic absorption lines as are present in the Earth's atmosphere (e.g. in the spectra of planets). This might also be true for gaseous disks around young stellar objects. To circumvent this, Bailey et al. (ref. [11]) propose a forward modeling approach in which one generates a model spectrum of the object, adds the effects of telluric transmission (both at significantly higher resolution than the actual observations), convolves it with a Gaussian or the numeric ILP to the actual resolution, and compares the result with the observed spectrum. v) Finally, it is possible that the chosen standard star has a faint, unknown intrinsic line that will influence the final spectrum. If very high precision is required, this effect has to be considered.

To illustrate this last point, we show in Fig. 3 the residuals of a K-band spectrum of the B8III star HIP 21949 observed with CRIRES. The observed spectrum was modeled with a synthetic spectrum generated by the RFM code, and the model spectrum was subtracted from the observed one. To eliminate instrumental features as much as possible, the same was done with the spectrum of another standard star (the one of Fig. 2), and its residuals were subtracted from the residuals of HIP 21949. This technique is similar to the one used by ref. [13], see next section. An emission feature at the level of roughly 1.4% of the continuum flux remains in these residuals. It is also present when dividing the two standard star spectra by each other, albeit with more noise. If this star was used as a standard star for a science target, this feature would mimic an absorption line in the science spectrum which is actually not present. HIP 21949 is known to be chemically peculiar[14], with large overabundances of Si. Since it is probably also a radial velocity variable with unknown period, we did not succeed to identify this feature with any atomic species. Without good knowledge of the NIR spectrum of a telluric standard star, such unknown features could compromise the science results.

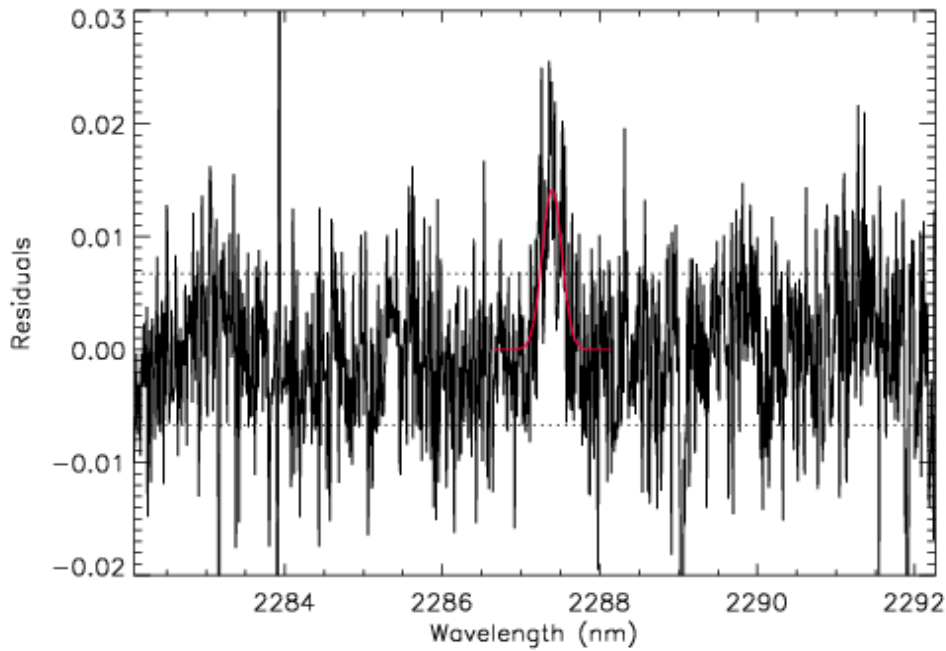

Fig. 3: Residuals of the CRIRES spectrum of HIP 21949 showing an unidentified feature in this star. Included as red graph is a Gaussian fit to the feature. The central wavelength of the Gaussian curve is 2287.40 ± 0.02 nm, and the width (sigma) is 0.13 nm. The dotted horizontal line shows the standard deviation of the residuals. Excluding the unidentified feature, the residuals have a standard deviation of about 0.5%, corresponding to a spectral fidelity of 1:200.

## 5.1 Observing standard star spectra with an IFU

Despite its disadvantages, the method of using standard star spectra described above is still widely in use because of its simplicity, and it seems that model spectra of telluric transmission will never fully replace standard star observations for complex and low-SNR science spectra. The high-resolution IFU of METIS with huge multiplexing advantages needs efficient calibration, hence a procedure to observe standard stars needs to be established. For extended science targets, it is obviously much too time consuming and cumbersome to observe a standard star spectrum in every single spaxel. Most importantly, the spectrum of a quasi-point source will not be adequate for correcting spectra of an extended source, because the slit illumination is different and thus the resolution (ILP) will be different. How to proceed from here?

Correcting a point source science spectrum for telluric lines is relatively straightforward. The standard star needs to be observed in exactly the same spaxel as the science target. The correction of the spectrum for telluric absorption lines can be achieved with the procedure described above. The goodness of the telluric correction will profit from stable seeing conditions and AO performance, as well as stable thermal conditions and water content of the atmosphere (as mentioned, small differences in airmass can be corrected for unsaturated lines). In view of the water content and its variability as well as thermal conditions, METIS will, probably like no other E-ELT instrument, profit from a high, dry, and as stable as possible observing site.

In case of an extended science target, the light of a telluric standard is required to evenly illuminate the IFU. This can be accomplished in different ways: The AO tip-tilt mirror can be used to steer a defocused standard star in a random pattern across the field to uniformly cover the IFU during an exposure; a random signal can be applied to the DM to defocus the standard star image even more than already done by atmospheric turbulence; a defocused standard star image can be moved across the IFU in steps in subsequent exposures. This latter method is further discussed in the following, but the other procedures may work in a very similar way.

The telescope is defocused such that the standard star "disk" covers a substantial part of the IFU. Subsequently, spectra are taken of this defocused standard star. The extended emission of the defocused standard star takes care automatically of the about 10% reduced spectral resolution caused by the even illumination of the slices, effectively dividing out the telluric lines. Finally, for every spectrum from every single spaxel, the above described procedure has to be followed for telluric correction. Ideally, this procedure is implemented in the instrument data reduction pipeline.

The current METIS IFU design foresees a 0.43×1.49 arcsecond FOV. This is a rather elongated shape. If the science target fills the whole IFU, this means that the standard star is best defocused to a radius of 0.36 arcseconds, and the IFU is then covered by three consecutive observations in which the defocused image is moved along the IFU in steps (with the AO switched off). Defocusing the standard star such that it fills the whole IFU in once is not recommended, since the flux will be substantially reduced, decreasing the SNR that is reached within a given time. Also, if the science target does not fill the whole IFU, the standard star might be less defocused or fewer steps may be required to acquire sufficient calibration data.

We used the METIS sensitivity code described elsewhere in this conference[15] to estimate the efficiency of the METIS IFS, i.e. the SNR of a typical early-type star with a given flux as a function of exposure time. Assuming a star defocused to cover one square-arcsecond and a Paranal-like atmosphere for the background radiation, the minimum continuum flux density at 4 μm for an SNR of 100 in 5 minutes exposure time is calculated to be 7 Jy, approximately a $4^{th}$ (L-band) magnitude star. A search on Simbad has been conducted to check how many stars can be found in the sky that are brighter than this limit. Since no whole-sky catalog in the L- or M-band is available, we used the approximation that all suitable standard stars have a spectrum similar to that of Vega ($0^{th}$ magnitude at every wavelength), hence should have a (K-L) color very close to 0.0. With this assumption we can easily search on Simbad for 2MASS K-band magnitudes brighter than 4.0. At a declination $\delta \leq +45$ degrees, corresponding to an airmass of $\leq 2.85$ viewed from the selected E-ELT site Cerro Armazones in Chile, 207 stars of spectral type A9 or earlier are found. There also exist at least 125 asteroids which reach a flux of 1 Jy or more at 5 μm at some point of their orbit (T. Müller, private communication, 2009). Hence, we conclude that enough calibrators are available in the sky to obtain observed transmission spectra with moderate SNR within a reasonably short exposure time. For high-SNR spectra, the number of standard stars is more limited.

## 6. COMBINING MODEL SPECTRA AND STANDARD STAR OBSERVATIONS

Another possibility is to combine the two approaches for telluric line correction, but it is challenging to get the best of both "worlds". One option here is to actually fit a model of telluric transmission to the the *standard* star spectrum, and to then divide the *science* spectrum by this model. This method is of benefit if the science spectrum has many intrinsic lines and cannot be fitted directly. Of course, by this approach one would fit the atmospheric parameters and spectral resolution to the standard spectrum, and not to the science spectrum itself. At least this approach would not add further noise to the final spectrum.

An interesting method that combines observed and modeled spectra of telluric transmission has been developed by A. Mandell at Goddard Space Flight Center[13]. In this method, a model is fit to both standard and science spectrum, and the two models are *subtracted* from the respective observed spectrum. Then the residuals of the science and the comparison star are differenced, removing remnant fringes and other instrumental artifacts. Thus, the method does not restore the science spectrum as it would have been observed in the absence of the Earth's atmosphere, but rather retrieves faint features that are present on top of the continuum emission of the science target. Flaws in the molecular line data or imprecise modeling will both cancel out with this procedure. However, any spurious features present in the standard star spectrum would also be added to the final residuals, see the warning above. This needs to be carefully checked for. The results in Mandell et al. reached the photon noise limit of the original spectra of SNR ≈ 2000. Promising results with this method have been achieved with medium resolution (R = 27,000) spectra taken with NIRSPEC at the Keck II telescope[13], and apparently good results have also been achieved for high-resolution CRIRES spectra (A. Mandell, private communication, 2010).

# 7. CONCLUSIONS

The novel design of the METIS IFS will require novel techniques for calibration to get the best out of the observed spectra. Correcting the absorption lines imposed by the Earth's atmosphere are one of the challenges to face in the calibration process. In this contribution, we explored different methods to correct for the telluric absorption lines in science spectra, and discussed their pros and cons.

The classical technique of division by a standard star spectrum will still be useful in the METIS era. It will probably mostly be applied on complex and low-SNR science spectra. However, it also has a number of disadvantages that can be overcome to a large part by model spectra of telluric transmission. The use of model spectra is an emerging technique with promising prospects. The atmospheric conditions at the time and in the direction of the science observations can be modeled precisely, and no further noise or spurious features will be added to the final spectrum. Nevertheless, a number of improvements have to be made to bring the model spectra to the highest level of accuracy. Molecular data need some improvement, and we propose here a collaboration with scientists working on the line lists (e.g. the HITRAN database). Also, the (minor) discrepancies between current codes to synthesize transmission spectra have to be sorted out. Modeling of the sky emission spectrum to determine the atmospheric conditions at the time of observation is another possibility that should be explored in more detail for the use with METIS spectra.

It is clear that model spectra will play a crucial role in the calibration process of METIS spectra, and they need to form a part of the instrument data reduction package. Only this will guarantee that the highest spectral fidelities will be achieved and that the METIS science requirements will be met.

**Acknowledgements** This research is based on observations at the Very Large Telescope of the European Southern Observatory, Cerro Paranal/Chile, under Programs 081.D-0669, 081.D-0702, 383.D-0894, and 179.C-0151. SU acknowledges support from the Fund for Scientific Research of Flanders (FWO) under grant number G.0470.07.